\documentclass[a4paper,11pt]{article}
\usepackage{pos}
\usepackage{natbib}  
\usepackage{hyperref} 
\usepackage{booktabs}

\title{Two-pole structures in QCD --
a universal phenomenon governed by chiral dynamics}

\author[a]{Jia-Ming Xie}
\author[a]{Jun-Xu Lu}
\author*[a,b,c,d,e]{Li-Sheng Geng}
\author[f,g,h,c,e]{Bing-Song Zou}

\affiliation[a]{School of Physics, Beihang University, Beijing 102206, China}

\affiliation[b]{
Sino-French Carbon Neutrality Research Center, \'Ecole Centrale de P\'ekin/School
of General Engineering, Beihang University, \\
Beijing 100191, China}

\affiliation[c]{Peng Huanwu Collaborative Center for Research and Education, Beihang University, \\
Beijing 100191, China}

\affiliation[d]{Beijing Key Laboratory of Advanced Nuclear Materials and Physics, Beihang University, \\
Beijing 102206, China }

\affiliation[e]{Southern Center for Nuclear-Science Theory (SCNT), Institute of Modern Physics, Chinese Academy of Sciences, \\
Huizhou 516000, Guangdong Province, China}

\affiliation[f]{Department of Physics, Tsinghua University, \\
Beĳing 100084, China}

\affiliation[g]{CAS Key Laboratory of Theoretical Physics, Institute of Theoretical Physics, Chinese Academy of Sciences, \\
Beijing 100190, China}

\affiliation[h]{School of Physical Sciences, University of Chinese Academy of Sciences, \\
Beijing 100049, China}

\emailAdd{ljxwohool@buaa.edu.cn}
\emailAdd{lisheng.geng@buaa.edu.cn}

\abstract{We illustrate how the two-pole structures of the $\Lambda(1405)$ emerge from the underlying universal chiral dynamics that describe the coupled-channel interactions between octet baryons and pseudo-Nambu-Goldstone bosons. Specifically, we attribute this phenomenon to the form of the leading-order chiral potential, which is of the Weinberg-Tomozawa type.  We reveal how the underlying chiral dynamics can be exposed by examining the light-quark mass evolution of the two poles. The latest lattice QCD simulations have indeed found evidence for the existence of the two poles of $\Lambda(1405)$, in qualitative agreement with our predictions. We briefly mention a recent work in which lattice QCD simulations are studied more quantitatively, along with a proposal for how the SU(3) flavor content of the two poles of $\Lambda(1405)$ can be experimentally verified. }

\FullConference{The 11th International Workshop on Chiral Dynamics (CD2024)\\
 26-30 August 2024\\
Ruhr University Bochum, Germany\\}

\begin{document}
\maketitle
Among numerous baryon excited states, the $\Lambda(1405)$, with quantum numbers $J^P=1/2^-$, $I=0$, and $S=-1$, is one of the most extensively studied baryon resonances. It is probably the first "exotic hadron" in the sense that it does not fit into the traditional constituent quark model~\cite{Capstick:1986ter} due to its significantly lighter mass compared to its nucleon counterpart, $N^*(1535)$, despite containing a heavier strange quark. Besides, the mass difference between $\Lambda(1405)$ and its spin partner $\Lambda(1520)$ with $J^P=3/2^-$ is significantly larger than the corresponding splitting in the nucleon sector, i.e., that between $N(1535)$ and $N(1520)$~\cite{Hyodo:2011ur}. Notably, despite the long-standing acceptance of its spin-parity being $1/2^-$, the experimental confirmation was achieved only recently, as detailed in Ref.~\cite{CLAS:2014tbc}.
  
The exotic nature of the $\Lambda(1405)$ has attracted considerable attention. Actually, even before its discovery~\cite{Hemingway:1984pz}, it was predicted theoretically to be a $\bar{K}N$ bound state based on the $\bar{K}N$ scattering data~\cite{Dalitz:1959dn,Kim:1965zzd}.
With the advancement of chiral unitary approaches integrating SU(3)$_L$$\times$SU(3)$_R$ and chiral dynamics and elastic unitarity~\cite{Kaiser:1995eg,Kaiser:1996js,Oset:1997it,Oller:2000ma,Hyodo:2011ur,Oller:2019opk,Mai:2020ltx}, the $\bar{K}N$ molecular interpretation of the $\Lambda(1405)$ has been further strengthened. An unforeseen outcome of these studies is that the $\Lambda(1405)$ corresponds to two dynamically generated poles on the second Riemann sheet of the complex energy plane~\cite{Oller:2000fj,Jido:2003cb}. These two poles lie between the thresholds of $\pi \Sigma$ and $\bar{K}N$. A recent unified description of meson-baryon scattering at the next-to-next-to-leading order (NNLO)~\cite{Lu:2022hwm} further confirms its two-pole structure. 

In previous studies, the origin of such two-pole structures has been attributed to (broken) SU(3) symmetry from the perspective of the SU(3) group. In  Ref.~\cite{Jido:2003cb}, it was demonstrated that in the SU(3) flavor symmetry limit, three bound states are anticipated: one singlet and two degenerate octets. In reality, where SU(3) symmetry is broken, the singlet evolves into the lower pole of the $\Lambda(1405)$, while one octet evolves into the higher pole.  Analogous findings have been reported for $K_1(1270)$~\cite{Roca:2005nm} and $D_0^*(2300)$~\cite{Kolomeitsev:2003ac,Guo:2006fu,Guo:2009ct,Albaladejo:2016lbb,Guo:2018tjx,Du:2020pui}. For a comprehensive review of two-pole structures from this perspective, we refer to Ref.~\cite{Meissner:2020khl}. However, like the Wigner-Eckart theorem~\cite{Sakurai:2011zz}, group theory alone does not fully elucidate the phenomenon. Specifically, it fails to reveal the underlying dynamics.

To render our discussion more focused, it is essential to define the “two-pole structures”.
In this talk, we define two-pole structures as two dynamically generated states: one resonant and one bound, with respect to the most strongly coupled channels. The two poles ought to be located near each other between two coupled channels and have a mass difference smaller than the sum of their widths. Consequently, the two poles overlap in the invariant mass distribution of their decay products, creating the perception of a single state. It should be noted that since both coupled channels are relevant, the manifestation of the state can vary across different reactions, depending on the specific reaction mechanism. 
According to this definition, the  $K_1(1270)$~\cite{Roca:2005nm,Geng:2006yb} and $\Xi(1890)$~\cite{Molina:2023uko} also qualify for two-pole structures. They are governed by the same chiral dynamics highlighted in the present work.

In the talk, the two-pole structures emerging from the underlying coupled-channel chiral dynamics and the pseudo Nambu-Goldstone~(pNG) nature of the pseudoscalar mesons are explored from the following three aspects, that is, whether the off-diagonal coupling between the two dominant channels play a decisive role, how the explicit chiral symmetry breaking generates the two-pole structures and whether the energy dependence of the Weinberg-Tomozawa~(WT) potential is relevant.

\section{Theoretical Framework}

The LO chiral Lagrangian describing the pseudoscalar-baryon~(PB) interaction is~\cite{Jido:2003cb}:
\begin{equation}
    \mathcal{L}_{\mathrm{PB}}^{\mathrm{WT}}=\frac{1}{4f^2}\mathrm{Tr}\left(\Bar{\mathcal{B}}i\gamma^{\mu}\left[\Phi\partial_{\mu}\Phi-\partial_{\mu}\Phi\Phi,\mathcal{B}\right]\right),
\end{equation}
from which one can obtain the $S$-wave potential in the center of mass (c.m.) frame
\begin{equation}
    V^{\mathrm{PB}}_{ij}=-\frac{C_{ij}}{4f^2}\left(2\sqrt{s}-M_{i}-M_{j}\right)=-\frac{C_{ij}}{4f^2}\left(E_i+E_j\right),
    \label{VPB}
\end{equation}
where the subscripts $i$ and $j$ represent the incoming and outgoing channels in the isospin basis, $M$ is the mass of the baryon, $E$ is the energy of the pseudoscalar meson, and $C_{ij}$ are the corresponding Clebsch-Gordan (CG) coefficients for the $\{\bar{K}N, \pi \Sigma \}$ coupled-channel system, which are listed in Table~\ref{tab:Cij1405}.
\begin{table}[htpb]
    \centering
    \renewcommand{\arraystretch}{1.6} 
    \setlength{\tabcolsep}{6pt} 
    \begin{tabular}{ccc}
        \toprule
         & $\bar{K}N$ & $\pi \Sigma$ \\
        \midrule
        $\bar{K}N$ & 3 & $-\sqrt{\frac{3}{2}}$ \\
        $\pi \Sigma$ & $-\sqrt{\frac{3}{2}}$ & 4 \\
        \bottomrule
    \end{tabular}
    \caption{CG coefficients for the   $\left\{\bar{K}N, \pi \Sigma \right\}$ coupled-channel system.}
    \label{tab:Cij1405}
\end{table}

The LO chiral Lagrangian for the pseudoscalar-vector~(PV) interaction is~\cite{Birse:1996hd,Roca:2005nm} 
\begin{equation}
    \mathcal{L}_{\mathrm{PV}}^{\mathrm{WT}}=-\frac{1}{4f^2}\mathrm{Tr}\left(\left[\mathcal{V}^{\mu},\partial^{\nu}\mathcal{V}_{\mu}\right]\left[\Phi,\partial_{\nu}\Phi\right]\right),
\end{equation}
from which one can  obtain the following potential projected onto $S$-wave,
\begin{equation}
    V^{\mathrm{PV}}_{ij}\left(s\right)=-\epsilon^i \cdot \epsilon^j \frac{C_{ij}}{8f^2}\left[ {3s-\left(M_i^2+m_i^2+M_j^2+m_j^2\right)}  {-\frac{1}{s}\left(M_i^2-m_i^2\right)\left(M_j^2-m_j^2\right)} \right],
    \label{VPV}
\end{equation}
where $M_{i,j}$  are the vector meson masses, $m_{i,j}$ are the pseudoscalar meson ones, and $\epsilon_{i,j}$ are the polarization vectors of the incoming and outgoing vector mesons. Close to the threshold, treating the light masses of the pseudoscalar mesons as small quantities and taking the chiral limit of $M_i=M_j\equiv M$, Eq.~(\ref{VPV}) can be simplified to
\begin{equation}
     V^{\mathrm{PV}}_{ij}\left(s\right)=-\epsilon^i \cdot \epsilon^j \frac{C_{ij}}{8f^2}4M\left(E_i+E_j\right),
\end{equation}
which is the same as Eq.~(\ref{VPB}) up to the scalar product of polarization vectors, trivial dimensional factors, and CG coefficients given in Table~\ref{tab:Cij1270}.  This similarity already hints that the same chiral dynamics govern the two systems. 

\begin{table}[htpb]
    \centering
    \renewcommand{\arraystretch}{1.6} 
    \setlength{\tabcolsep}{6pt} 
    \begin{tabular}{ccc}
        \toprule
         & $K^*\pi$ & $\rho K$ \\
        \midrule
        $K^*\pi$ & -2 & $\frac{1}{2}$ \\
        $\rho K$ & $\frac{1}{2}$ & -2 \\
        \bottomrule
    \end{tabular}
    \caption{CG coefficients for the  $\left\{K^*\pi, \rho K \right\}$ coupled-channel system.}
    \label{tab:Cij1270}
\end{table}

It turns out that the above leading order~(LO) chiral Lagrangians are responsible for generating the two-pole structures of $\Lambda(1405)$ and $K_1(1270)$. Higher-order potentials do not appear to significantly modify the picture~\cite{Ikeda:2012au,Guo:2012vv,Mai:2012dt,Lu:2022hwm,Zhou:2014ila,Zhuang:2024udv}. 

The unitarized amplitude reads~\cite{Oller:2000ma}, 
 \begin{equation}
    T=\left(1-VG\right)^{-1}V,
    \label{BSeq}
\end{equation}
where $G$ is a diagonal matrix with elements $G_{kk}\equiv G_{k}\left(\sqrt{s}\right)$.  The loop function $G_k\left(\sqrt{s}\right)$ of channel $k$ is logarithmically divergent and can be regulated either in the dimensional regularization (DR) scheme or the cutoff scheme. In the DR scheme, a subtraction constant is introduced, whereas in the cutoff scheme, a cutoff is required. In practice, the subtraction constants or cutoff values are determined by fitting to the scattering data and should be of natural size~\cite{Oller:2000ma,Roca:2005nm}.

The PB loop function, regularized in the dimensional regularization scheme, reads~\cite{Oller:2000fj} \begin{equation} \begin{aligned}
     G^{\mathrm{PB}}_k\left(\sqrt{s}; a_k\left(\mu\right)\right)=&\frac{2M_k}{16\pi^2} \left\{ {a_k\left(\mu\right)+\frac{m_k^2-M_k^2+s}{2s}\mathrm{ln}\frac{m_k^2}{M_k^2}} \right. \\& \left. {+\mathrm{ln}\frac{M_k^2}{\mu^2}+\frac{p_k\left(\sqrt{s}\right)}{\sqrt{s}}\mathrm{ln}\left(\frac{\phi_{++}\phi_{+-}}{\phi_{-+}\phi_{--}}\right)}\right\},
     \end{aligned}
 \end{equation}
 where
 \begin{equation}
     \phi_{\pm \pm} \equiv \pm s\pm \left(M_k^2-m_k^2\right)+2\sqrt{s}p_k\left(\sqrt{s}\right),
 \end{equation}
 and $a_k\left(\mu\right)$ is the subtraction constant.

The  PV loop function~\cite{Roca:2005nm} has a similar form 
\begin{equation}
G^{\mathrm{PV}}_k(\sqrt{s}; a_k\left(\mu\right)) = \left(1+\frac{1}{3}\frac{p_k^2}{M_k^2}\right) \frac{1}{2M_k}G^{\mathrm{PB}}_k\left(\sqrt{s}; a_k\left(\mu\right)\right),
\end{equation}
along with the modification of the Bethe-Salpeter equation~\ref{BSeq} as follows
\begin{equation}
T^{\mathrm{PV}}=\left(1+V^{\mathrm{PV}}G^{\mathrm{PV}}\right)^{-1}\left(-V^{\mathrm{PV}}\right)\epsilon^i \cdot \epsilon^j.
\end{equation}

One can quantify the magnitude of the couplings of a resonance/bound state to its constituents using the residues of the corresponding pole on the complex energy plane, which can be calculated in the following way,
\begin{equation}
    g_i g_j=\lim_{\sqrt{s} \rightarrow z_R}\left(\sqrt{s}-z_R\right)T_{ij}\left(\sqrt{s}\right),
    \label{coupling}
\end{equation}
where $z_R\equiv m_R-i\Gamma_R/2$ is the pole position.

\begin{figure*}[htpb]
    \centering
    \includegraphics[width=3.6in]{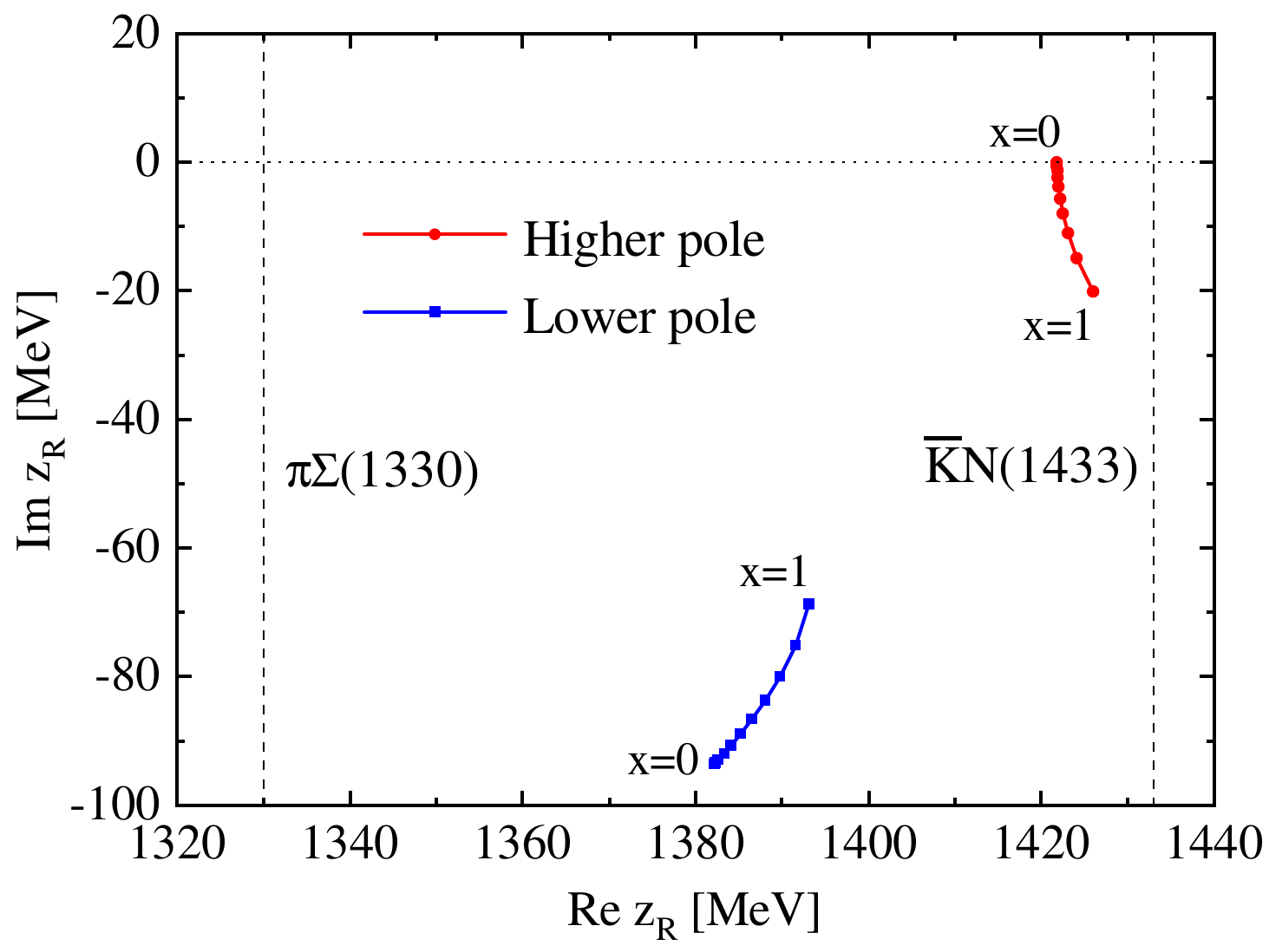}
    \caption{Evolution of the two poles of $\Lambda\left(1405\right)$ as a function of the off-diagonal potential $x \times V_{\bar{K}N-\pi\Sigma}$ with $0\le x\le 1$. Every point on the lines is taken in steps of $x=0.1$.}
    \label{fig:trajectory}
\end{figure*}

\section{Results and Discussions}
In the following, we try to answer the above three questions one by one. For this, we use the $\Lambda(1405)$ as an explicit example.
\subsection{Q1: Coupled channel effects}\label{Q1}

We focus on the two most relevant channels, namely $\bar{K}N$ and $\pi\Sigma$, in the isospin $0$ and strangeness $-1$ sector around the 1400 MeV region~\cite{Jido:2003cb,Hyodo:2007jq}, while neglecting the $\eta\Lambda$ and $K\Xi$ channels. Two poles are generated, \textit{i.e.}, $W_H=1426.0-20.1i$ MeV and $W_L=1393.1-68.7i$ MeV, with the following subtraction constants $a_{\bar{K}N}=-1.95$ and $a_{\pi \Sigma}=-1.92$. These pole positions are consistent with the LO~\cite{Jido:2003cb}, next-to-leading order~(NLO)~\cite{Ikeda:2012au,Guo:2012vv,Mai:2012dt}, and NNLO results~\cite{Lu:2022hwm}.
\begin{figure*}[htpb]
    \centering
    \includegraphics[width=3.6in]{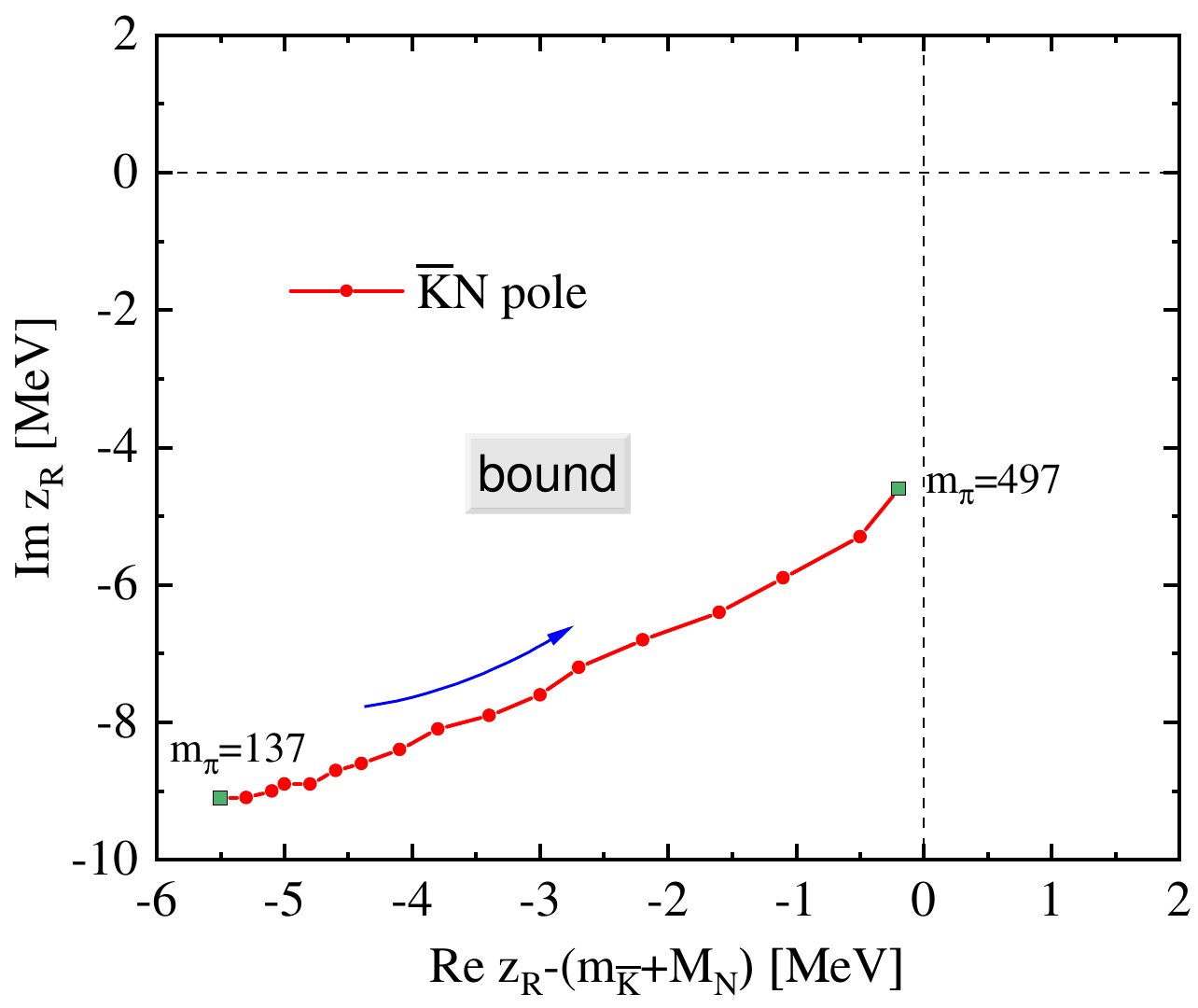}
       \includegraphics[width=3.6in]{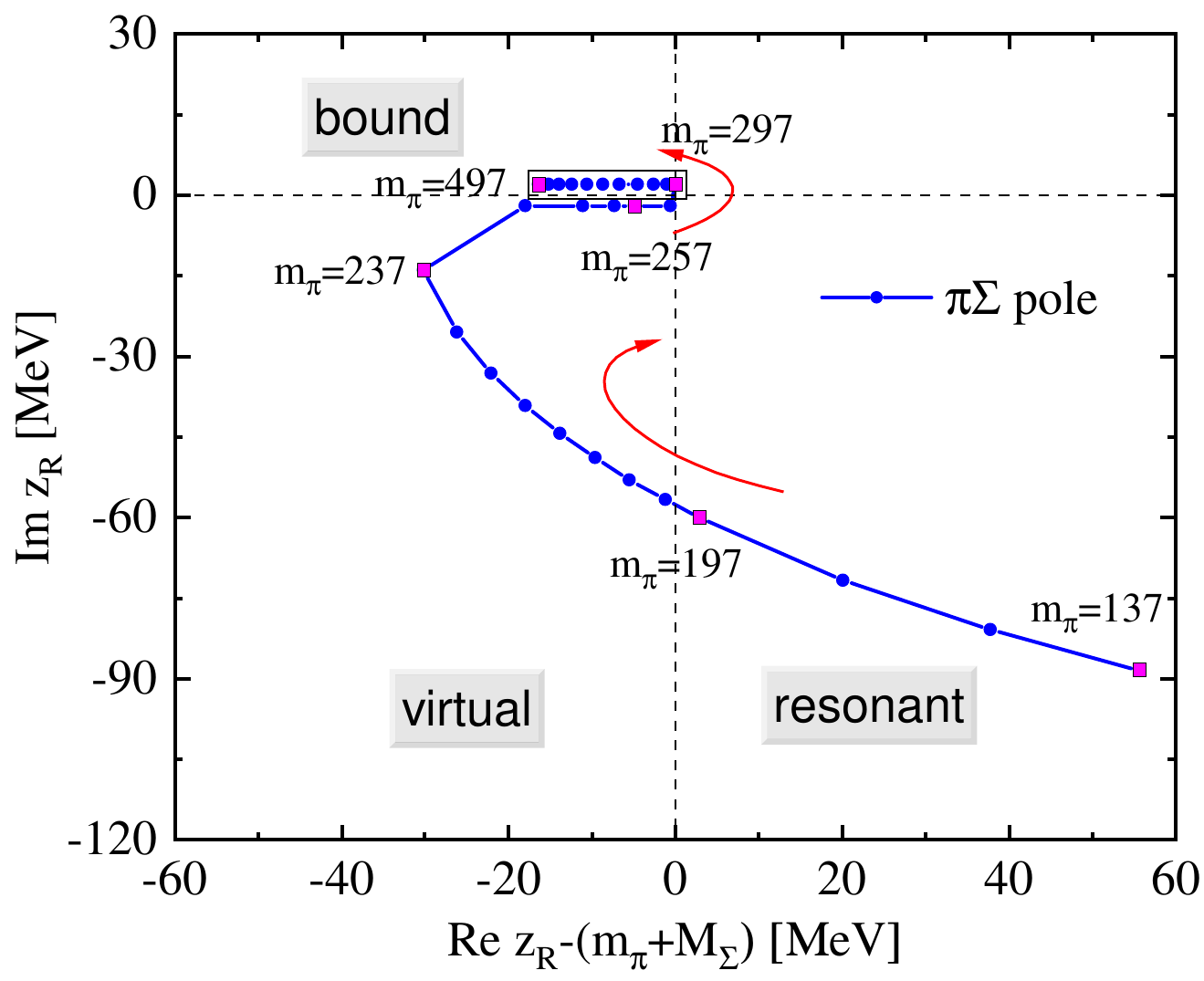}
    \caption{Trajectories of the  two poles of $\Lambda(1405)$ as functions of the pion mass  $m_{\pi}$ from 137 MeV to 497 MeV. Critical masses are labeled by solid squares, between which the points are equally spaced.}
    \label{fig:2channel}
\end{figure*}

Naively, one expects that the two poles are linked to the coupling between the $\bar{K}N$ and $\pi\Sigma$ channels, i.e., coupled channel effects. However, this is not the case. To demonstrate this, we multiply the off-diagonal matrix elements of the WT potential by a factor $x$, where $0\le x\le1$, and then obtain the evolution of the two poles, as depicted in Fig.~\ref{fig:trajectory}. It is evident that even in the case of complete decoupling, i.e., $x = 0$, the two poles still emerge between the $\Bar{K}N$ and $\pi\Sigma$ thresholds. Meanwhile, due to the lack of phase space, the imaginary part of the higher pole approaches zero.  The coupling between the two channels causes the two poles to shift to higher energies. This enables the higher pole to decay into the $\pi \Sigma$ channel, thereby acquiring a finite width. Nevertheless, it is crucial to emphasize that the coupling between the two channels is not the main factor responsible for the emergence of two dynamically generated states between the two relevant channels. The significance of the coupling lies in the fact that without it, the higher pole would not be observable in the invariant mass distribution of the lower $\pi\Sigma$ channel.

\subsection{Q2: Explicit chiral symmetry breaking}\label{Q2}

As is well known in QCD, chiral symmetry is broken in two distinct manners: spontaneously and explicitly. The role of explicit symmetry breaking can be appreciated by studying the pole trajectories as a function of the light-quark (pion) mass. Since the coupling between $\bar{K}N$ and $\pi\Sigma$ does not play a decisive role in the formation of the two-pole structure, we focus on single channels.

According to Eq.~(\ref{VPB}), the diagonal WT interaction is proportional to the energy of the pseudoscalar meson. They read
\begin{equation}
\begin{aligned}
    V_{\Bar{K}N- \Bar{K}N}\left(\sqrt{s}\right)=&-\frac{6}{4f^2}E_{\Bar{K}}=-\frac{6}{4f^2}\sqrt{m_{\Bar{K}}^2+q_{\Bar{K}}^2},\\
    V_{\pi \Sigma- \pi \Sigma}\left(\sqrt{s}\right)=&-\frac{8}{4f^2}E_{\pi}=-\frac{8}{4f^2}\sqrt{m_\pi^2+q_{\pi}^2}.
    \label{V-diagonal}
\end{aligned}
\end{equation}

Owing to the explicit chiral symmetry breaking, the mass of the kaon is significantly larger than that of the pion. Therefore, the $\bar{K}N$ interaction is stronger than the $\pi \Sigma$ one near the threshold, which leads to a $\bar{K}N$ bound state. In addition, the energy dependence, combined with the small mass of the pion, amplifies the $q^2$ term of the $\pi\Sigma$ interaction.  Consequently, this amplification is responsible for the existence of a $\pi\Sigma$ resonance. 

As the pion mass changes, the masses of the baryons and the kaon also vary accordingly. At leading order and for a physical strange quark mass, $m_K^2=a+bm_{\pi}^2$, where $a=0.291751$ GeV$^2$ and $b=0.670652$~\cite{Ren:2012aj}. One can
then obtain the eta mass, following the Gell-Mann–Okubo mass formula~\cite{Gell-Mann:1961omu,Okubo:1961jc,Okubo:1962zzc}, i.e., 
\begin{equation}
    m_{\eta}^2 = \frac{4m_{K}^2-m_{\pi}^2}{3}.
\end{equation}

For the baryon octet, we employ the covariant baryon chiral perturbation theory (ChPT) to describe their light-quark mass dependence. Up to $\mathcal{O}\left(p^2\right)$, the octet baryon masses receive contributions from both the chiral limit and the tree level. They read 
\begin{equation}
M_B\left(m_{\pi}\right)=M_0+M_B^{\left(2\right)}=M_0+\sum_{\phi=\pi, K}\xi_{B, \phi}m_{\phi}^2,
\label{mBdep}
\end{equation}
where $M_0$ is the chiral limit baryon mass and $\xi_{B, \phi}$ are the relevant coefficients that contain three low-energy constants $b_0$, $b_D$, and $b_F$, shown in Table 1 of Ref.~\cite{Ren:2012aj}. The mass dependence relation, Eq.~\ref{mBdep}, is fitted to the lattice QCD data of the PACS-CS Collaboration~\cite{PACS-CS:2008bkb}, as presented in Table 8 of Ref.~\cite{Ren:2012aj}. The low energy constants are shown below in Table~\ref{tab:parameters}~\cite{Song:2018qqm}.

\begin{table}[htpb]
    \centering
    \renewcommand{\arraystretch}{1.2} 
    \setlength{\tabcolsep}{3pt} 
    \begin{tabular}{cccc}
        \toprule
        $M_0$~[MeV] & $b_0$~[GeV$^{-1}$] & $b_D$~[GeV$^{-1}$] & $b_F$~[GeV$^{-1}$] \\
        \midrule
        962.3 & -0.22809 & 0.038183 & -0.15768 \\
        \bottomrule
    \end{tabular}
    \caption{Chiral limit baryon mass $M_0$, and three low-energy constants $b_0$, $b_D$, and $b_F$.}
    \label{tab:parameters}
\end{table}

With the light-quark mass dependence of all relevant quantities fixed as described above, we can solve the BS equation and obtain the trajectories of the two poles of $\Lambda(1405)$, as shown in Fig.~\ref{fig:2channel}. The evolution of the higher pole is relatively simple. As the pion mass increases, both its real and imaginary parts decrease. Note that as the pion mass increases, the two thresholds also increase in value. On the other hand, the trajectory of the lower pole is highly nontrivial. As the pion mass increases, it first becomes a virtual state from a resonant state for a pion mass of approximately 200 MeV, then becomes a virtual state for a pion mass of around 300 MeV, and remains so up to a pion mass of approximately 500 MeV.

\subsection{Q3: Energy dependence of the WT potential}\label{Q3}
To emphasize the importance of the energy dependence of the chiral potential in shaping the two-pole structure, we replace $E_i+E_j$ in the chiral potential of Eq.~(\ref{VPB}) with $m_i+m_j$. With the original subtraction constants, we obtain only one pole at $1413.3-13.2i$ MeV, corresponding to a $\bar{K}N$ bound state. 
 We verified that switching off the off-diagonal interaction has little effect on our conclusion. As the pion mass is much smaller than the kaon mass, the attraction of the $\pi \Sigma$ single channel is weaker than that of the $\Bar{K}N$ single channel and thus cannot support a bound state. Certainly, if we increase the strength of the attractive potential, we can obtain two bound states, but not a bound state and a resonant state, and as a result, there is no longer a two-pole structure. 

\section{Recent lattice QCD and theoretical studies}
Such a non-trivial dependence of the pole positions on the light-quark masses has been verified by a recent lattice QCD simulation~\cite{BaryonScatteringBaSc:2023zvt,BaryonScatteringBaSc:2023ori}.
The simulations are performed for unphysical quark masses,  $m_{\pi} \approx 200$ MeV and $m_K \approx 487$ MeV. The obtained scattering amplitudes reveal the presence of a virtual bound state below the $\pi\Sigma$ threshold, strongly coupled to the $\pi\Sigma$ channel, in addition to a resonance pole near the $\bar{K}N$ threshold with a dominant coupling to the $\bar{K}N$ channel. By employing different K-matrix parameterizations, all fits to the lattice QCD results support the two-pole structure, consistent with the predictions from SU(3) chiral symmetry and unitarization. Note that the trajectories shown in Fig.~\ref{fig:2channel} rely on the meson and baryon masses as functions of the masses of the light quarks $u$, $d$, and $s$, or of the light mesons $\pi$ and $K$.  They are determined by fitting to the lattice QCD data of the PACS-CS Collaboration~\cite{PACS-CS:2008bkb,Ren:2012aj,Song:2018qqm} with the NLO baryon chiral perturbation theory. The recent lattice QCD study~\cite{BaryonScatteringBaSc:2023zvt,BaryonScatteringBaSc:2023ori}, on the other hand, was performed for $m_{\pi}\approx 200$ MeV and $m_K\approx 487$ MeV with the CLS D200 configuration. Although the descriptions of the properties of the two poles are similar to our discussion of the higher and lower pole trajectories in Section \ref{Q2}, they differ due to the distinct light-quark mass dependence of the physical quantities involved.

After the lattice QCD studies~\cite{BaryonScatteringBaSc:2023zvt,BaryonScatteringBaSc:2023ori}, many theoretical studies followed~\cite{Zhuang:2024udv,Ren:2024frr,He:2024uau}. In particular, it was shown in Ref.~\cite{Zhuang:2024udv} the results obtained in the lattice QCD simulations~\cite{BaryonScatteringBaSc:2023zvt,BaryonScatteringBaSc:2023ori} can be well described by the NLO unitarized ChPT using the quark mass dependence of the octet baryon masses~\cite{RQCD:2022xux} for the CLS ensembles and the NLO meson masses and decay constants determined in Ref.~\cite{Molina:2020qpw}. By extrapolating the lattice QCD results to the physical point using the one-loop NLO chiral Lagrangian under the framework of covariant baryon chiral perturbation theory, they confirmed the presence of two poles: a lower pole that is a virtual bound state and a higher pole located just below the $\bar{K}N$ threshold. They also predicted the behavior of these poles along different symmetry-breaking trajectories, including the SU(3) symmetric limit.

\section{More two-pole structures}
 From the above analysis, it is evident that replacing the matter particles (ground-state baryons) with ground-state vector mesons may also lead to the existence of two-pole structures. This is indeed the case as shown in Refs.~\cite{Roca:2005nm,Geng:2006yb}, where $K_1(1270)$ is found to correspond to two poles. The most relevant channels are  $K^* \pi(1030)$ and  $\rho K(1271)$, while the other three channels ($\omega K(1278)$, $K^* \eta(1440)$, and $\phi K(1515)$) play a negligible role. 
In Ref.~\cite{Geng:2006yb}, with $\mu=900$ MeV,   $a\left(\mu\right)=-1.85$, and $f=115$ MeV,
where $\mu$ is the renormalization scale, $a\left(\mu\right)$ is the common subtraction constant, and $f$ is the pion decay constant, one finds two poles located at  $W_H=1269.3-1.9i$ MeV and  $W_L=1198.1-125.2i$ MeV below the $\rho K$ and above the $K^* \pi$ thresholds.  Eliminating the three higher channels, one can obtain almost the same two poles located at  $W_H=1269.5-12.0i$ MeV and $W_L=1198.5-123.2i$ MeV,
by slightly adjusting the subtraction constants as $a_{K^{\ast}\pi}=-2.21$ and $a_{\rho K}=-2.44$.

In Refs.~\cite{Molina:2023uko,MolinaPeralta:2024ihc}, Molina \textit{et al.} investigated the nature of the $\Xi(1820)$ resonance using the chiral unitary approach in the $\Sigma^*\bar{K}$-$\Xi^*\pi$-$\Xi^*\eta$-$\Omega K$ coupled-channel system. The study proposes that $\Xi(1820)$ is not a single resonance but rather consists of two distinct states—one narrow and one broad. This theoretical prediction agrees with the recent BESIII measurement of the $K^-\Lambda$ mass distribution in the $\psi(3686)$ decay to $K^-\Lambda \bar{\Xi}^+$, which observed a $\Xi(1820)$ width significantly larger than the Particle Data Group (PDG) average. Using the interaction of the pseudoscalar mesons with the decuplet baryons, the authors derived the coupled-channel scattering amplitudes. They solved the BS equation to identify two poles associated with $\Xi(1820)$. They further demonstrated that these two states naturally explain the large width observed in the BESIII experiment, which could not be accounted for by a single-resonance interpretation. Further studies on the two-pole structure of $\Xi(1820)$ were performed in Refs.~\cite{Liang:2024fsv,Duan:2024ygq}, in which two different final-state   $\Xi^*\pi$ and $\bar{K}^0\Sigma^{*-}$ invariant mass distribution were predicted in the $\psi(3686)$ weak decay. By analyzing the mass distributions and invariant mass spectra, the authors compared their theoretical results with the experimental findings. Their results confirmed that the broader state contributes significantly to the higher-mass region, while the narrower state remains consistent with the previous PDG estimates.

\begin{figure*}[htpb]
    \centering
    \includegraphics[width=3.6in]{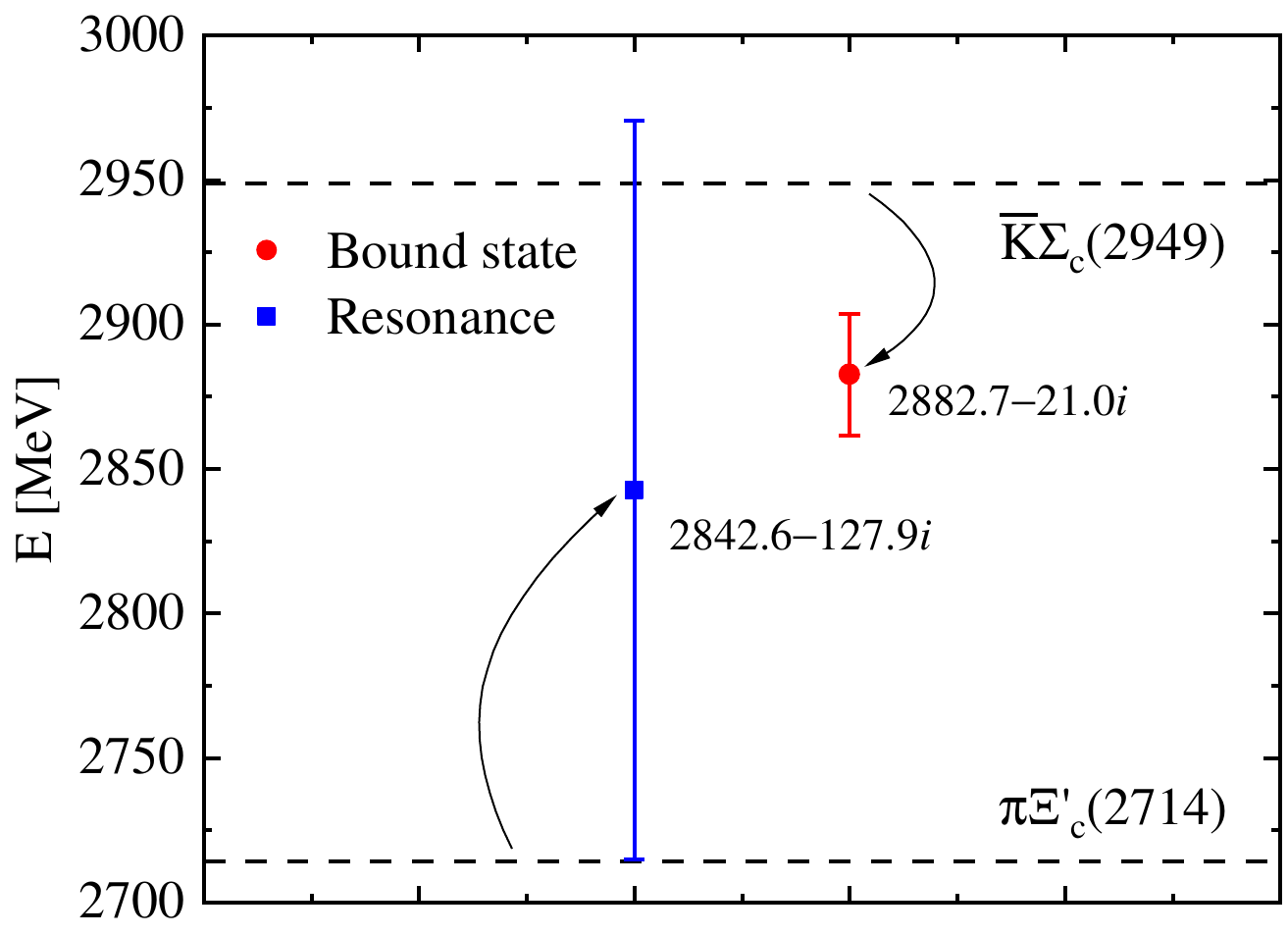}
    \caption{Two poles of the $\Bar{K}\Sigma_c-\pi \Xi_c'$ system. 
    The arrows denote the dominant channels. The vertical bars are the widths corresponding to twice the imaginary parts of the pole positions.}
    \label{fig:single charm baryon}
\end{figure*}

Given the universality of chiral dynamics discussed above and inspired by the aforementioned two-pole structures $\Lambda(1405)$, $K_1(1270)$, and $\Xi(1820)$ appearing in the interactions between various heavy matter particles and pseudoscalar mesons, more such two-pole structures in other systems composed of a pair of heavy matter particles and pseudoscalar mesons are foreseeable, such as the singly charmed baryon sector~\cite{Lu:2014ina}. Using the criteria proposed in this work, one can identify two channels $\Bar{K}\Sigma_c(2949)$ and $\pi\Xi_c'(2714)$ that can generate such two-pole structures. 

\begin{table}[htpb]
    \centering
    \renewcommand{\arraystretch}{1.5} 
    \setlength{\tabcolsep}{6pt} 
    \begin{tabular}{ccc}
        \toprule
         & $\Bar{K}\Sigma_c(2949)$ & $\pi\Xi_c'(2714)$ \\
        \midrule
        $\Bar{K}\Sigma_c(2949)$ & 3 & $-\frac{1}{\sqrt{2}}$ \\
        $\pi\Xi_c'(2714)$ & $-\frac{1}{\sqrt{2}}$ & 2 \\
        \bottomrule
    \end{tabular}
    \caption{CG coefficients for the  $\left\{\bar{K}\Sigma_c, \pi \Xi_c' \right\}$ coupled-channel system.}
    \label{tab:Cijcharm}
\end{table}

The interactions between charmed baryons and pseudoscalar mesons are the same as the PB interaction in Eq.~\ref{VPB}, which reads.
\begin{equation}
V^{\bar{K}\Sigma_c-\pi\Xi_c'}_{ij}\left(\sqrt{s}\right) = -\frac{C_{ij}}{4 f^2} (E_i + E_j),
\end{equation}
where the CG coefficients $C_{ij}$ are shown in Table~\ref{tab:Cijcharm}. With a common cutoff of $\Lambda=800$ MeV for both coupled channels, one finds two poles in the isospin $1/2$ channel, one at $W_H=2882.7-21.0i$ MeV and the other at $W_L=2842.6-127.9i$ MeV. The higher pole strongly couples to the $\Bar{K}\Sigma_c$  channel, while the lower pole couples more to the $\pi\Xi_c'$ channel, as shown in Fig.~\ref{fig:single charm baryon}. 

\begin{figure*}[htpb]
    \centering
    \includegraphics[width=3.6in]{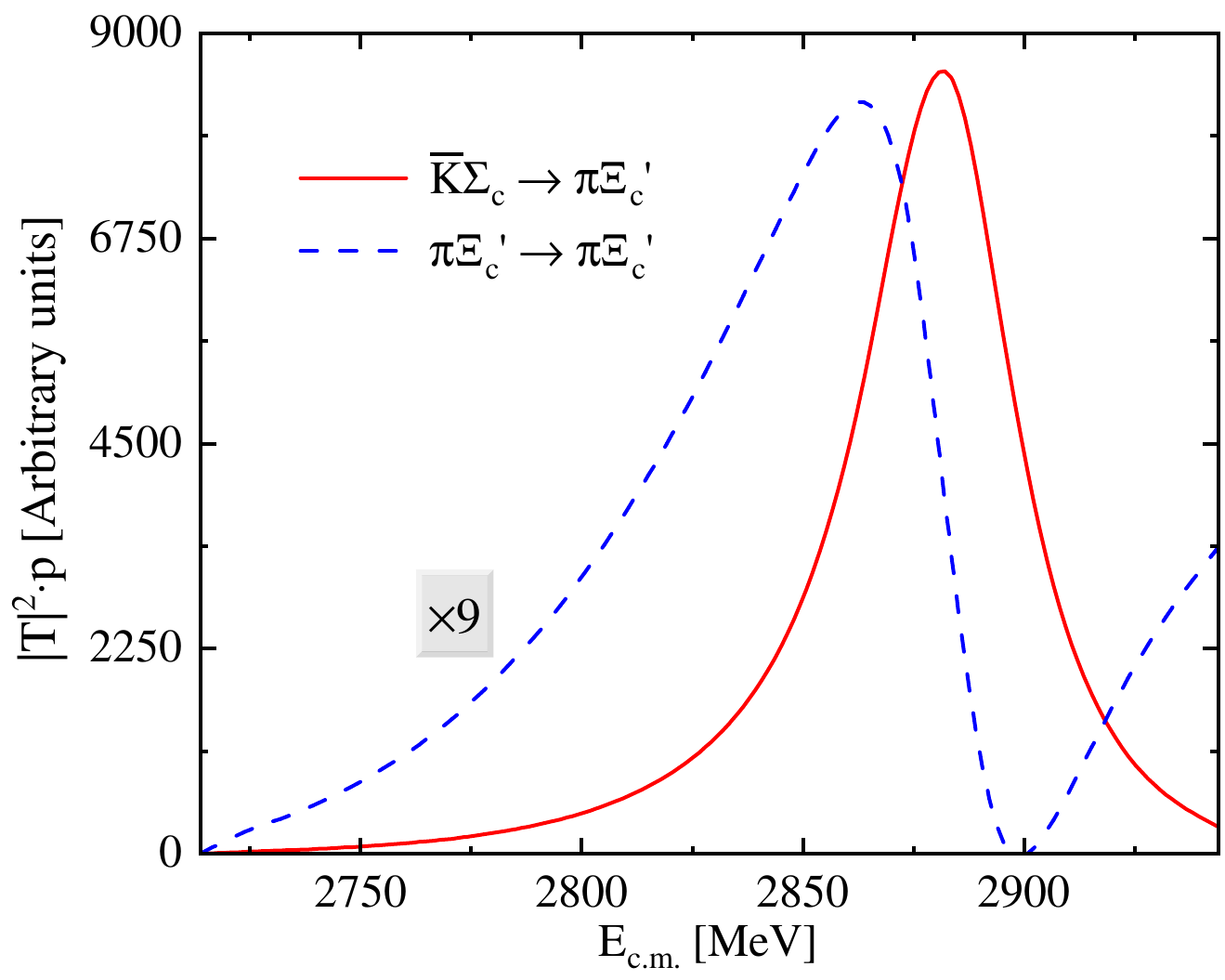}
    \caption{Invariant mass distributions of $\pi \Xi_c'$  (in arbitrary units) as functions of the c.m. energy,  $\left|T_{\Bar{K}\Sigma_c \rightarrow \pi\Xi_c'}\right|^2 p_{\pi}$ (red solid) and $9\times \left|T_{\pi\Xi_c' \rightarrow \pi\Xi_c'}\right|^2 p_{\pi}$ (blue dashed), where $p_{\pi}$ is the 3-momentum of the pion in the c.m. frame of the final states.}
    \label{fig:double peak}
\end{figure*}

Following Ref.~\cite{Lu:2014ina}, with the unitarized amplitudes $\Bar{K}\Sigma_c \rightarrow \pi\Xi_c'$ and $\pi\Xi_c' \rightarrow \pi\Xi_c'$, one can construct the $\pi \Xi_c'$ invariant mass distributions shown in Fig.~\ref{fig:double peak}. It should be noted that the lineshapes of the $\Bar{K}\Sigma_c \rightarrow \pi\Xi_c'$ and $\pi\Xi_c' \rightarrow \pi\Xi_c'$ exhibit significant overlap. However, they peak at slightly different positions and have different widths. This can be qualitatively accounted for by the fact that the $\Bar{K}\Sigma_c \rightarrow \pi\Xi_c'$ receives a more significant contribution from the higher pole, while the $\pi\Xi_c' \rightarrow \pi\Xi_c'$ coupling is more pronounced for the lower pole. We need to emphasize that, despite the two-pole structure being intrinsically related to the underlying chiral dynamics, regularization—specifically, the cutoff—plays a crucial role. Consequently, we advocate for further theoretical and experimental investigations of the predicted two-pole structures.

\section{Further experimental verifications}
In Ref.~\cite{Jido:2003cb}, the LO chiral potential predicted that the lower pole of the $ \Lambda(1405)$ is predominantly an SU(3) singlet, while the higher pole is predominantly an SU(3) octet. This was confirmed in Ref.~\cite{Zhuang:2024udv} with the NLO potential.
On the other hand, Ref.~\cite{Guo:2023wes} predicted that at NLO, the SU(3) flavor contents of the two poles are exchanged. 
Recently, Ref.~\cite{He:2024uau} proposed a novel method to identify the two-pole structure of the $\Lambda(1405)$, more precisely, their SU(3) flavor contents, in the hadronic decays of charmonia into $\bar{\Lambda}\Sigma\pi$ and $\bar{\Lambda}(1520)\Sigma\pi$. The argument is that charmonia and $\bar{\Lambda}(1520)$  are SU(3) flavor singlets, but the $\bar{\Lambda}$ belongs to the SU(3) octet.  It was demonstrated that this SU(3) flavor filter remains effective even when considering flavor symmetry breaking. Therefore, with the vast charmonium data sets collected, one can gain a better understanding of the $\Lambda(1405)$.

\section{Summary and outlook}
We have demonstrated how the chiral dynamics contained in the Weinberg-Tomozawa potential is
responsible for generating two-pole structures that are relevant to understanding $\Lambda(1405)$, $K_1(1270)$, and
$\Xi(1820)$. Explicit chiral symmetry breaking leads to a kaon mass larger than that of the pion; together with the energy dependence of the chiral potential, they collaboratively generate two poles between two nearby coupled channels. One should note that the degeneracy of the two coupled channels is also lifted by explicit chiral symmetry breaking. In particular, we point out that the evolution of the two poles as a function of light quark masses can help verify the underlying chiral dynamics. Recent lattice QCD simulations~\cite{BaryonScatteringBaSc:2023zvt,BaryonScatteringBaSc:2023ori} and theoretical studies~\cite{Zhuang:2024udv,Ren:2024frr,He:2024uau} do seem to support such an idea. Further experimental studies, such as those proposed in Ref.~\cite{He:2024uau}, can help further advance our understanding of the nature of this mysterious phenomenon. 

In this work, we have not discussed the widely studied $D_0^*(2300)$~\cite{Kolomeitsev:2003ac,Guo:2006fu,Guo:2009ct,Albaladejo:2016lbb,Guo:2018tjx,Du:2020pui}.  because the two states theoretically related to it are well separated and, therefore, do not qualify as two-pole structures, as defined here. See the Supplemental Material of Ref.~\cite{Xie:2023cej} for more discussions about this mysterious state.  One should note
that there are other works discussing two-pole structures, such as Refs.~\cite{Zhou:2020moj,Clymton:2024pql,Karliner:2024cql,Cui:2025wfo}, but we refrain from discussing these works because the underlying mechanism differs from ours.
\bibliographystyle{JHEP}
\bibliography{ref}



\end{document}